\documentclass[12pt]{elsarticle}
\usepackage{amsmath, amssymb, amsthm}
\usepackage{geometry, graphicx, multicol, float}
\newcommand{\rev}[1]{#1}
\begin{document}

\begin{frontmatter}

\title{A Saturation-Based Optimal Velocity Model for Traffic Flow Dynamics}

\author[umich]{Nizhum Rahman}
\author[umich]{Trachette L. Jackson}

\address[umich]{Department of Mathematics, University of Michigan, Ann Arbor, MI 48109, United States}

\begin{abstract}
Many headway-based car-following models describe longitudinal adaptation through linear relaxation laws, which can produce unrealistically large accelerations and limit the physical consistency of microscopic traffic dynamics. Motivated by this limitation, we develop a saturation-based extension of the classical Optimal Velocity Model (OVM) that preserves the headway-dependent desired-speed structure while introducing bounded nonlinear acceleration dynamics. Linear stability analysis shows that the proposed formulation preserves the classical long-wave instability mechanism associated with stop-and-go waves while modifying the stability threshold and enforcing bounded acceleration. Ring-road simulations support the analysis and illustrate how the model alters perturbation growth, wave amplitude, and relaxation behavior relative to the classical OVM. The resulting framework provides a compact and analytically tractable extension for studying nonlinear traffic-wave dynamics and physically constrained car-following behavior.
\end{abstract}



\end{frontmatter}


\section{Introduction}

\rev{Microscopic traffic modeling has gained renewed importance because connected and automated vehicles, vehicle-to-everything (V2X) communication, and data-driven control strategies require reliable descriptions of how vehicles accelerate, decelerate, and interact \cite{Han2022V2X,Wang2023Review,Lach2024Review,Rowan2025Review}.}
Traffic flow exhibits complex collective behavior such as stop-and-go waves, spontaneous traffic jams, and flow breakdowns arising from interactions among individual vehicles. At the microscopic level, these effects are commonly modeled through \emph{car-following} laws, which describe how drivers adjust their motion in response to surrounding traffic conditions.

One of the most influential microscopic models is the Optimal Velocity Model (OVM), introduced by Bando et al.~\cite{Bando1995}. Despite its simplicity, the OVM reproduces key traffic phenomena, including traveling waves, traffic jams, and nonlinear instabilities, making it a foundational model in traffic-flow research across physics, engineering, and applied mathematics \cite{Helbing2001,Nagatani2002,Kerner2004,Wilson2011}.  
\rev{Following standard traffic-flow terminology, the expression ``optimal velocity'' refers here to a headway-dependent desired speed rather than to global optimality among all existing traffic models.}

\rev{The continuing relevance of the OVM is reflected in recent extensions incorporating safety constraints \cite{Abdelhalim2022,Fei2023}, stochastic delayed dynamics \cite{Meng2023}, and distributed control strategies for connected vehicle platoons \cite{Zhu2025}. Representative related models also include formulations with heterogeneous driver classes and cooperative throttle-based control \cite{Zhai2021TimidAggressive,Zhai2022Throttle}. These developments demonstrate that the key challenge is no longer the historical relevance of the OVM itself, but how its analytical transparency can be adapted to modern requirements of safety, automation, and physically realistic vehicle response.}

From the standpoint of nonlinear dynamics, the OVM supports traveling waves, metastable regimes, and soliton-like structures~\cite{Komatsu1995}.  
\rev{Linear stability analysis and the associated dispersion relation explain how equilibrium traffic loses stability, while experiments such as the circular-track study of Sugiyama et al.~\cite{Sugiyama2008} demonstrate that stop-and-go waves can emerge spontaneously even without bottlenecks.}
These results underscore the value of the OVM both as a practical traffic model and as a prototype for nonlinear traffic-wave phenomena \cite{Orosz2010,Treiber2013}.

\medskip
\noindent\textbf{Gap and motivation.}
\rev{Despite its success, a fundamental limitation remains in the classical OVM formulation.}
Most OVM-type models still assume a \emph{linear relaxation law},
\[
    \dot v \propto V(\Delta x)-v,
\]
which is mathematically convenient but can imply unrealistically large accelerations.  
\rev{Recent reviews emphasize that modern microscopic traffic models must increasingly balance interpretability, automation relevance, data compatibility, and physical realism \cite{Han2022V2X,Wang2023Review,Lach2024Review,Rowan2025Review}. At the same time, the dominant direction of recent OVM development has focused on safety, uncertainty, stochasticity, delay, and control design \cite{Abdelhalim2022,Fei2023,Meng2023,Li2023,Zhu2025}, while the explicit incorporation of a drag-based saturation mechanism into the OVM acceleration law itself remains comparatively unexplored.}
\rev{Real vehicles accelerate under powertrain limits and aerodynamic resistance, producing bounded and smooth approaches to equilibrium rather than purely linear relaxation. Such physically grounded saturation mechanisms are rarely incorporated directly into headway-based car-following laws \cite{Wilson2008,Seibold2013}.}
Bounded acceleration is closely related to driver comfort, fuel efficiency, and automated-vehicle control \cite{Milanes2014,Kesting2010,Qin2024}.

\medskip
\noindent\textbf{Contribution of this work.}
To address this gap, we propose a \emph{Hybrid Optimal Velocity--Drag (OVD) model}.  
\rev{The proposed model preserves the OVM principle that desired speed depends on headway, but replaces the classical linear relaxation law with a drag-based saturation mechanism that enforces bounded acceleration and smooth convergence toward equilibrium.}
\rev{The resulting formulation introduces a physically grounded acceleration mechanism within the OVM framework, providing a physically grounded extension of classical relaxation-based car-following dynamics while preserving analytical tractability and the classical instability structure associated with stop-and-go waves.}

\medskip
\noindent\textbf{Main contributions.}
\rev{The present work provides a physics-informed extension of the classical OVM framework that complements modern approaches centered on safety, control, stochasticity, and automation.}

\begin{itemize}
    \item \rev{We introduce a hybrid OVD model that embeds drag-limited acceleration saturation within the classical OVM framework while preserving its headway-based behavioral structure.}
    
    \item \rev{We derive the corresponding linearized dynamics, dispersion relation, and long-wave stability threshold, showing explicitly how the drag term modifies the classical OVM stability criterion.}
    
    \item \rev{We demonstrate analytically and numerically that the model preserves the instability mechanisms associated with stop-and-go waves while eliminating the unrealistic unbounded accelerations produced by the classical relaxation law.}
\end{itemize}

\rev{Natural quantitative comparisons between the proposed model and existing car-following laws include maximal admissible acceleration, the linear stability threshold, perturbation growth or decay rates, and stop-and-go wave amplitude.}

This distinguishes the Hybrid OVD model from several existing formulations:

\begin{itemize}
    \item Unlike the Full Velocity Difference Model (FVDM)~\cite{Jiang2001}, which stabilizes traffic through velocity-difference feedback, the Hybrid OVD model leaves the interaction law purely headway-based and instead modifies the acceleration dynamics themselves.

    \item Engineering-oriented drag-based formulations such as Fadhloun \& Rakha~\cite{Fadhloun2020} incorporate aerodynamic resistance and powertrain effects, but do not directly connect these dynamics with OVM-style wave instability theory.
\end{itemize}

The resulting model preserves the nonlinear instability structure of the OVM while enforcing bounded acceleration.

\medskip
\noindent\textbf{Scope and relation to existing models.}
The Hybrid OVD model is not intended as a replacement for velocity-difference extensions such as the FVDM~\cite{Jiang2001} or multi-anticipative models~\cite{Treiber2000}. Rather, it isolates the \emph{physical} saturation mechanism associated with drag and bounded acceleration.  
\rev{In this sense, the present study should be viewed as a baseline physics-oriented extension that can later be combined with richer information channels, delay terms, or control-theoretic components.}
\rev{Because the present paper focuses on analytical model development, calibration against public trajectory datasets and broader empirical validation are left for future work rather than treated as prerequisites for the present mechanism and stability study.}

\medskip
\noindent\textbf{Overview of the paper.}
We first review the classical OVM and its stability properties.  
\rev{We then introduce the Hybrid OVD model, derive the corresponding dispersion relation and long-wave stability threshold, and state the principal stability results in theorem form.}
\rev{The numerical section reports the parameter choices used in simulation and illustrates stable and unstable regimes through the comparison indicators described above.}
We conclude with implications, applications, and possible extensions of the hybrid framework.

\section{Classical Optimal Velocity Model}

The Optimal Velocity Model (OVM), introduced by Bando et al.~\cite{Bando1995}, is one of the most widely studied microscopic car-following models. In this framework, each driver adjusts their acceleration based on the difference between their current speed and a desired speed that depends on the headway to the vehicle in front.

Consider $N$ vehicles moving on a ring road of length $L$. Let $x_n(t)$ and $v_n(t)$ denote the position and velocity of the $n$-th vehicle at time $t$. The headway is
\[
\Delta x_n(t) \;=\; x_{n+1}(t) - x_n(t),
\]
with periodic indexing $x_{N+1} = x_1 + L$, and $\Delta x_n(t) > 0$ denotes the spacing between successive vehicles.

The OVM equations of motion are
\begin{equation}
\label{eq:ovm}
    \frac{dx_n}{dt} = v_n, 
    \qquad
    \frac{dv_n}{dt} = \alpha \big(V(\Delta x_n) - v_n\big),
\end{equation}
where $\alpha > 0$ is a constant sensitivity parameter and 
\rev{$V:\mathbb{R}^+ \to \mathbb{R}^+$ is a monotonically increasing function, commonly referred to as the optimal velocity function in the traffic-flow literature.}

This formulation represents relaxation toward a headway-dependent desired velocity and serves as a standard baseline for microscopic traffic modeling.

\medskip
\noindent\textbf{Classical stability criterion.}
A key feature of the OVM is that the uniform-flow equilibrium
\(
v_n = V(b),\ \Delta x_n = b
\)
can lose stability and generate stop-and-go waves. Linearizing
\eqref{eq:ovm} shows that long-wave perturbations grow when
\[
\alpha\,V'(b) \;>\; \frac{1}{2c},
\qquad c = V(b),
\]
while smaller $\alpha$ or shallower slopes $V'(b)$ yield a stable
uniform flow \cite{Bando1995,Helbing2001,Treiber2013}.

\medskip
\noindent\textbf{Common choices for the optimal velocity function.}
While the OVM formalism only requires that $V(\Delta x)$ be increasing
and bounded, many studies use smooth sigmoidal forms such as
\[
V(\Delta x) = \frac{v_{\max}}{2}\Big[
\tanh\!\big( (\Delta x - h_c)/\ell \big) + 1
\Big],
\]
or variants of the original hyperbolic tangent law proposed in
\cite{Bando1995}. These choices determine the slope $V'(b)$ governing stability. Other formulations introduce cutoffs to allow $V(\Delta x)=0$ for small gaps \cite{Ge2004}.

\medskip
\noindent\textbf{Limitations of the classical formulation.}
The linear relaxation law implies unbounded acceleration:
\[
\dot v_n = \alpha\big(V(\Delta x_n) - v_n\big),
\]
so large speed differences lead to unrealistically large accelerations. This discrepancy is particularly relevant in dense traffic and transient regimes, motivating the drag-based saturation mechanism introduced in the Hybrid OVD model.

\medskip
\noindent\textbf{Relation to the Hybrid OVD model.}
The Hybrid OVD formulation retains the OVM structure but replaces the linear relaxation term with a nonlinear saturation law. Near equilibrium, the hybrid model reduces locally to
\(
\dot v \approx \alpha_{\mathrm{eff}}(V - v),
\)
with a state-dependent sensitivity, preserving the OVM as the underlying behavioral framework.

\section{Hybrid OVD model}

The Optimal Velocity Model (OVM) captures the idea that drivers adjust their acceleration 
according to the difference between their actual velocity and a headway-dependent target 
velocity. While elegant, the OVM uses a \emph{linear relaxation law}, which means that the 
larger the difference $V(\Delta x_n)-v_n$, the stronger the acceleration. This can yield 
unrealistically large accelerations and does not directly reflect vehicle dynamics. The 
Hybrid Optimal Velocity--Drag (OVD) model introduced in this section modifies this 
assumption by incorporating a physically motivated nonlinear saturation mechanism.

\subsection{Physical motivation from drag-limited motion}

In physics, a single vehicle subject to a constant forward force from the engine and a 
quadratic aerodynamic drag force satisfies
\begin{equation}
\label{eq:drag}
m \dot v = F_{\text{eng}} - k v^2,
\end{equation}
where $m$ is the vehicle’s mass, $F_{\text{eng}}$ is the effective forward force generated 
by the engine, and $k v^2$ represents aerodynamic drag.

\medskip

Defining
\[
a := \frac{F_{\text{eng}}}{m}, 
\qquad 
v_{\max} := \sqrt{\frac{F_{\text{eng}}}{k}},
\]
we obtain
\begin{equation}
\label{eq:3}
\dot v = a\left(1 - \frac{v^2}{v_{\max}^2}\right).
\end{equation}

This law has two key properties:
\begin{itemize}
    \item At low speeds, $\dot v \approx a$.
    \item As $v\to v_{\max}$, acceleration smoothly decays to zero.
\end{itemize}

\medskip
\noindent\textbf{Interpretation of the parameter \(a\).}
The quantity
\[
a = \frac{F_{\text{eng}}}{m}
\]
represents the low-speed acceleration capability of the vehicle. Typical passenger vehicles 
yield values $a \approx 1.5\text{--}4.0~\mathrm{m/s^2}$, consistent with standard vehicle 
dynamics references~\cite{Rajamani2012,Guzzella2013}. In the Hybrid OVD model, \(a\) thus 
provides a physically interpretable parameter for calibration.

\medskip
\noindent\textbf{Motivation for the hybrid model.}
In dense traffic, the equilibrium speed is determined by headway rather than an intrinsic 
vehicle property. This motivates replacing the fixed $v_{\max}$ with the desired speed 
$V(\Delta x_n)$.

\subsection{Formulation of the hybrid model}

To combine behavioral realism with physical saturation, we replace $v_{\max}$ in the drag 
law by $V(\Delta x_n)$. The resulting Hybrid Optimal Velocity--Drag (OVD) model is
\begin{equation}
\label{eq:hybrid}
\frac{dx_n}{dt} = v_n, 
\qquad
\frac{dv_n}{dt} = a\left(1 - \frac{v_n^2}{V(\Delta x_n)^2}\right),
\end{equation}
where $a>0$ and $V(\Delta x_n)$ is a monotonically increasing, bounded function.

In contrast to the OVM, the effective sensitivity 
\[
\alpha_{\mathrm{eff}}(v,V)=\frac{a(V+v)}{V^2}
\]
depends on the current state.

\subsection{Interpretation of the dynamics}

Equation~\eqref{eq:hybrid}, consisting of the position update and the velocity dynamics, 
admits three intuitive regimes:

\begin{itemize}
    \item \textbf{Small headways:} When $\Delta x_n$ is small, the desired speed 
    $V(\Delta x_n)$ is close to zero, and the velocity equation becomes strongly 
    negative. This yields smooth, nonlinear deceleration and reduces the risk of 
    collisions.
    
    \item \textbf{Large headways:} When $\Delta x_n$ is large, $V(\Delta x_n)\approx 
    v_{\max}$ (the free-flow speed), and Eq.~\eqref{eq:hybrid} reduces to  
    \[
    \dot v = a\!\left(1-\frac{v^2}{v_{\max}^2}\right),
    \]
    i.e., the classical drag-limited law, giving smooth convergence to free-flow speed.
    
    \item \textbf{Intermediate headways:} For moderate gaps, drivers accelerate when 
    $v_n < V(\Delta x_n)$, but as $v_n$ approaches $V(\Delta x_n)$, the factor 
    $1-(v_n/V)^2$ gradually suppresses further acceleration. This produces smooth 
    transitions and eliminates unrealistically sharp responses.
\end{itemize}

\begin{figure}[H]
    \centering
\includegraphics[width=1.00\textwidth]{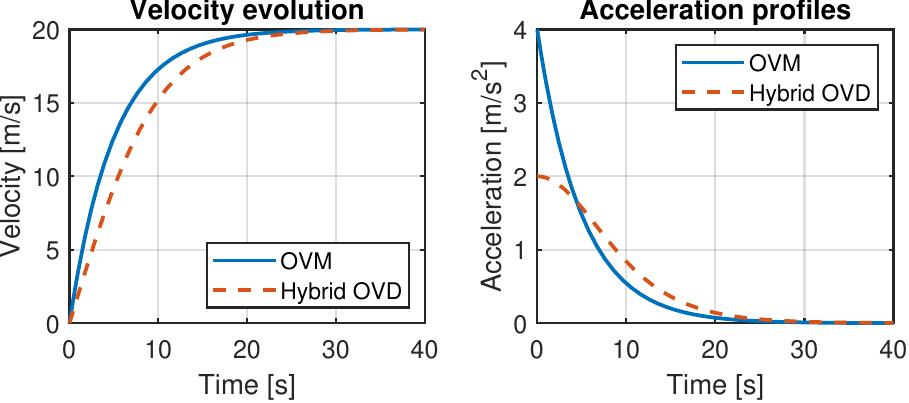}
    \caption{
    Single-vehicle response of the classical OVM and the Hybrid OVD model
    relaxing toward a prescribed constant target speed $V = 20~\mathrm{m/s}$.
    In this illustrative test, the OVM is given the acceleration law
    $\dot v = \alpha (V - v)$ with $\alpha = a / V = 0.1~\mathrm{s^{-1}}$,
    while the Hybrid OVD model uses $\dot v = a\bigl(1 - (v/V)^2\bigr)$
    with $a = 2.0~\mathrm{m/s^2}$. These choices ensure that both models
    have the same initial acceleration from rest, $\dot v(0) = a$.
    \textbf{Left:} velocity $v(t)$; both models converge to the same terminal
    speed $V$, but the OVM exhibits a purely linear relaxation, whereas the
    Hybrid OVD shows a drag-limited, sigmoidal approach.
    \textbf{Right:} acceleration $\dot v(t)$; the OVM produces relatively
    large accelerations when $v \ll V$, while the Hybrid OVD maintains
    acceleration bounded by $a$ and decays smoothly as $v \to V$, illustrating
    the physically motivated saturation in the Hybrid formulation.
    }
    \label{fig:singlecar-OVM-vs-Hybrid}
\end{figure}

These qualitative features are reflected in the single-vehicle dynamics shown in
Fig.~\ref{fig:singlecar-OVM-vs-Hybrid}. Even when both models relax toward the same
target speed and are initialized with the same acceleration from rest, the classical
OVM exhibits a linear, high-magnitude acceleration response, whereas the Hybrid OVD
model approaches the target speed smoothly with bounded, drag-limited acceleration.
\rev{In particular, the difference between the OVM and Hybrid OVD curves reflects how acceleration is regulated in the two models. In the OVM, acceleration is proportional to the speed difference, leading to relatively strong responses in the early stage (when $v \ll V$) and a linear decay toward equilibrium. In contrast, the Hybrid OVD model enforces a strict upper bound on acceleration and introduces a nonlinear, drag-like decay, resulting in a smoother and more gradual approach to the target speed. This distinction highlights how the Hybrid formulation avoids abrupt acceleration changes and provides a more physically realistic representation of vehicle dynamics.}
This example highlights the fundamental dynamical difference between the two formulations before we proceed to their analytical comparison.

\subsection{Relation to the Classical OVM}

Rewriting the hybrid velocity law,
\[
\dot v = a\left(1-\frac{v^2}{V^2}\right)
= \frac{a(V+v)}{V^2}(V-v),
\]
reveals that the structure resembles the OVM but with a state-dependent sensitivity
instead of a constant relaxation rate. Thus, the hybrid model maintains the behavioral 
principle of accelerating toward a desired speed while modifying the global acceleration 
profile to enforce physical realism.

\subsection{Local correspondence with the OVM}
\label{sec:local}

Although the Hybrid OVD model is nonlinear in general, it reduces to the classical OVM 
structure in the neighborhood of an equilibrium velocity. Let $v = V + \delta v$ with 
$|\delta v| \ll V$. Substituting into the hybrid law gives
\[
\dot v = \frac{a(V+v)}{V^2}(V-v)
= -\frac{a}{V^2}(2V+\delta v)\,\delta v.
\]
Expanding to leading order yields
\[
\dot v \approx \frac{2a}{V}\,(V - v),
\]
which is precisely an OVM relaxation law with an \emph{effective sensitivity}
\[
\alpha_{\text{eff}} = \frac{2a}{V}.
\]
Thus, small deviations from equilibrium are governed by a linear relaxation law, 
demonstrating that the hybrid model preserves the local structure of the OVM while 
enforcing globally bounded accelerations.

\begin{figure}[H]
    \centering
    \includegraphics[width=0.9\textwidth]{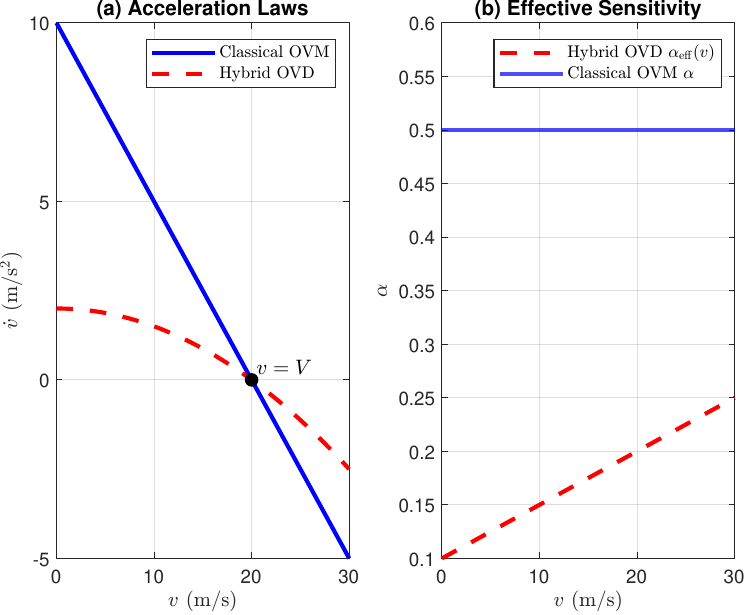}
    \caption{Comparison between the classical OVM and the Hybrid OVD 
    model. \textbf{(a)} Acceleration laws $\dot v$ versus velocity $v$
    for fixed $V=20$ m/s. \textbf{(b)} Effective sensitivity. The Hybrid
    model enforces bounded acceleration and a velocity-dependent 
    effective sensitivity, whereas the classical OVM uses a linear law 
    with constant sensitivity.}
    \label{fig:ovm-hybrid-comparison}
\end{figure}

To further compare the dynamical response of the two models, we examine the 
\emph{jerk}, defined as $j = d^2 v/dt^2$. For the classical OVM, the linear 
relaxation law $\dot v = \alpha (V - v)$ yields
\[
j_{\mathrm{OVM}} = -\alpha^2 (V - v),
\]
so the jerk grows linearly and without bound as $|v-V|$ increases. In contrast, the 
Hybrid OVD acceleration $\dot v = a\bigl(1 - v^2/V^2\bigr)$ gives
\[
j_{\mathrm{Hybrid}} 
= -\frac{2a^2}{V^2}\,v\left(1-\frac{v^2}{V^2}\right),
\]
a cubic law that vanishes at $v=0$ and $v=V$ and remains bounded for 
$0 \le v \le V$.

\begin{figure}[H]
    \centering
    \includegraphics[width=0.7\textwidth]{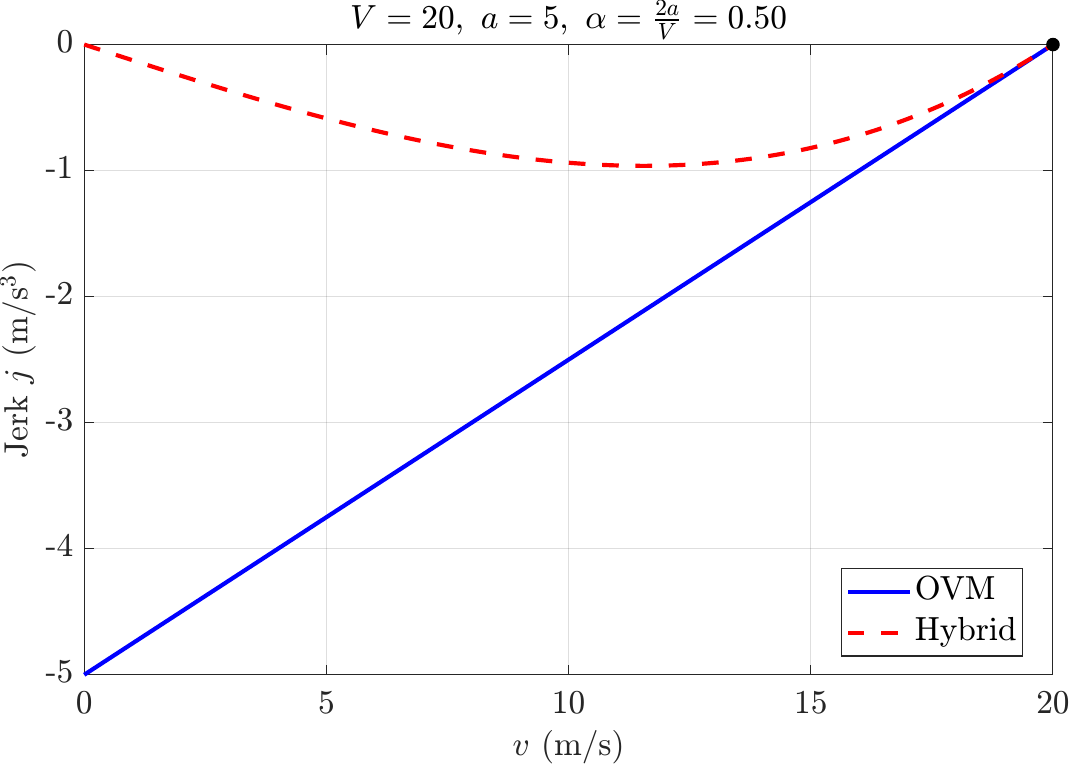}
    \caption{Jerk $j=d^2v/dt^2$ in the classical OVM and Hybrid OVD
    model for $V=20$~m/s, $a=5.0$~m/s$^2$, and matched sensitivity
    $\alpha = 2a/V = 0.5$~s$^{-1}$. The OVM exhibits a linear jerk law
    that grows without bound as $|v-V|$ increases. The Hybrid OVD model
    produces a cubic jerk law that remains bounded and vanishes at
    $v=0$ and $v=V$, but grows cubically for $v>V$.}
    \label{fig:ovm-hybrid-jerk}
\end{figure}

Figures~\ref{fig:ovm-hybrid-comparison} and~\ref{fig:ovm-hybrid-jerk} together 
highlight the physical interpretation of the Hybrid OVD model. While the classical 
OVM assumes a linear response with constant sensitivity and unbounded jerk, the 
hybrid formulation ensures that acceleration and jerk remain small and bounded in 
the physically relevant regime $0 \le v \le V$. Drivers accelerate more rapidly when far from equilibrium but naturally ease off as they approach their target speed, 
reflecting the influence of drag-limited dynamics. For overspeed states 
$v>V$, the Hybrid OVD jerk grows cubically—a modeling trade-off discussed further 
in Section~\ref{sec:limitations}. Detailed derivations of these jerk laws are provided 
in~\ref{app:jerk-analysis}.


\subsection{Comparison with classical car-following models}
\label{sec:model-comparison}

Before analyzing the linear stability properties of the Hybrid OVD
model, it is helpful to situate it relative to several major classes of
car–following laws: the classical Optimal Velocity Model (OVM), the
Full Velocity Difference Model (FVDM), the Intelligent Driver Model (IDM), 
and dynamics-aware engineering models such as the Fadhloun–Rakha (FR) 
formulation.  
Table~\ref{tab:model-comparison} summarizes several structural features
that distinguish these models from one another and from the proposed
Hybrid OVD law.

\begin{table}[H]
    \centering
    \small
    \begin{tabular}{lccccc}
        \hline
        Feature & OVM & FVDM & IDM & FR model & Hybrid OVD \\
        \hline
        Headway–only law                     & $\checkmark$ & $\times$ & $\times$ & $\times$ & $\checkmark$ \\
        Includes velocity-difference effects & $\times$     & $\checkmark$ & $\checkmark$ & $\times$ & $\times$ \\
        Includes vehicle dynamics / drag     & $\times$     & $\times$ & $\times$ & $\checkmark$ & $\checkmark$ \\
        Acceleration bounded by construction & $\times$     & $\times$ & $\checkmark$ & $\checkmark$ & $\checkmark$ \\
        Wave / instability theory available  & $\checkmark$ & $\checkmark$ & $\checkmark$ & $\times$ & $\checkmark$ \\
        \hline
    \end{tabular}
    \caption{Compact comparison of selected car–following models and the 
    proposed Hybrid OVD model.}
    \label{tab:model-comparison}
\end{table}

The OVM provides a foundational behavioral model whose simplicity admits
transparent dispersion relations and wave-instability theory, but its
linear relaxation law produces unrealistically unbounded acceleration.

The FVDM~\cite{Jiang2001} adds velocity-difference feedback, improving
string stability and reducing oscillatory amplification.  
Because it modifies the interaction law rather than the acceleration
mechanics, its purpose is complementary to the Hybrid OVD model: the
FVDM alters driver responsiveness, whereas the Hybrid OVD alters the
\emph{physical} saturation governing acceleration.

The IDM improves behavioral realism through its dependence on both 
headway and relative speed, and it enforces an implicit upper bound on
acceleration, but its free-road term is not derived from vehicle
dynamics and its analytical structure is more involved.

Engineering-oriented models such as the FR formulation incorporate
vehicle powertrain limits, drag forces, and heterogeneous driver inputs,
yielding physically realistic acceleration envelopes.  
Their complexity, however, makes them less amenable to closed-form
stability analysis or to a direct connection with classical traffic-wave
theory.

The Hybrid OVD model occupies a complementary niche: it retains the
OVM’s headway-based behavioral structure and preserves its tractable
instability mechanisms, while incorporating a simple and physically
interpretable drag-based saturation law that guarantees bounded
acceleration.  
In this sense, the Hybrid OVD provides a bridge between behaviorally
inspired models (OVM, FVDM, IDM) and physics-aware dynamics-based 
approaches (FR), offering both realism and analytical clarity.

This perspective motivates the stability analysis in the next section,
where we show that the Hybrid OVD model admits a clean dispersion
relation and a threshold condition that generalizes the classical OVM
criterion.

\section{Stability Analysis of the Hybrid OVD Model}

The Hybrid OVD model preserves the structure of classical car-following systems of the form
\[
\dot v_n = f(\Delta x_n, v_n, v_{n+1}),
\]
so standard tools of linear stability theory apply directly
\cite{Treiber2013}. In this section we restate the criteria for the OVD dynamics, derive the dispersion
relation, and connect the analytical threshold to numerical simulations.

\subsection{Model and steady (uniform-flow) solution}

We consider the Hybrid OVD system
\begin{equation}
\label{eq:hyb}
\frac{dx_n}{dt}=v_n, 
\qquad
\frac{dv_n}{dt}=a\!\left(1-\frac{v_n^2}{V(\Delta x_n)^2}\right),
\qquad
\Delta x_n := x_{n+1}-x_n,
\end{equation}
on a ring of length $L$ with $N$ vehicles ($x_{n+N}=x_n+L$).  

We seek a uniformly translating solution
\[
x_n^0(t)=nb+ct,\qquad v_n^0=c,
\]
with spacing $b=L/N$. Then $\Delta x_n^0=b$, and
\[
0=a\!\left(1-\frac{c^2}{V(b)^2}\right)\;\Rightarrow\; c=\pm V(b).
\]
Taking forward motion gives
\begin{equation}
\label{eq:uniform}
x_n^0(t)=nb+V(b)\,t,\qquad v_n^0=V(b).
\end{equation}

\subsection{\rev{Linearization about the uniform flow equilibrium}}

\rev{We linearize system~\eqref{eq:hyb} about the uniform flow equilibrium.}
Introduce perturbations
\[
x_n = x_n^0 + \xi_n,\qquad 
v_n = v_n^0 + \eta_n,
\quad
\Delta x_n = b + (\xi_{n+1} - \xi_n).
\]
Let $c := V(b)$ and $f := V'(b)$. Linearization of \eqref{eq:hyb} yields
\begin{align}
\dot{\xi}_n &= \eta_n, \label{eq:lin1}\\
\dot{\eta}_n &= -\frac{2a}{c}\,\eta_n 
              \;+\; \frac{2a}{c}\,f\,(\xi_{n+1} - \xi_n).
\label{eq:lin2}
\end{align}
\rev{Equations~\eqref{eq:lin1}--\eqref{eq:lin2} represent the linearized perturbation system governing small deviations from uniform flow.}
\rev{Here, the term $-2a/c$ reflects the drag-based relaxation, while the term involving $V'(b)$ represents coupling between neighboring vehicles.}

\subsection{Normal modes and dispersion relation}

Seeking normal modes
\[
\xi_n = \hat{\xi}\,e^{\lambda t + i k n},\qquad
\eta_n = \hat{\eta}\,e^{\lambda t + i k n},
\]
gives
\[
\lambda \hat{\xi} = \hat{\eta},\qquad
\lambda \hat{\eta} = -\frac{2a}{c}\,\hat{\eta} + \frac{2af}{c}\,(e^{ik} - 1)\hat{\xi}.
\]
Eliminating $\hat{\eta}$ yields
\begin{equation}
\boxed{\;
\lambda^2 + \frac{2a}{c}\,\lambda + \frac{2af}{c}\,(1 - e^{ik}) = 0,
\;}
\label{eq:dispersion-hybrid}
\end{equation}
\rev{which is obtained by eliminating the perturbation amplitudes from the linearized system. This equation relates the growth rate $\lambda$ of perturbations to the wave number $k$, and constitutes the dispersion relation of the Hybrid OVD model.}

\subsection{Long-wave expansion and stability threshold}
\label{sec:stability-longwave}

For $k\ll1$, use $e^{ik}=1+ik-\tfrac12k^2+O(k^3)$ and 
$\lambda = i\mu k + \nu k^2 + O(k^3)$ to obtain $\mu=f$ and
\[
\nu = \frac{c}{2a} f^2 - \frac{f}{2}.
\]
Thus
\[
\lambda(k)
= i f k
 + \Big( \frac{c}{2a} f^2 - \frac{f}{2} \Big) k^2
 + O(k^3).
\]

The leading real part is
\[
\Re \lambda \approx 
\Big( \frac{c}{2a} f^2 - \frac{f}{2} \Big) k^2,
\]
which is negative provided
\[
a > a^\ast, 
\qquad 
a^\ast := c\,V'(b).
\]

Hence uniform flow is
\begin{itemize}
    \item stable if $a > a^\ast$,
    \item unstable if $a < a^\ast$,
    \item neutrally stable if $a = a^\ast$.
\end{itemize}

This shows that drag-based saturation rescales the effective relaxation rate.

\medskip
\noindent
\emph{Connection to simulations.}
In Section~5, parameters satisfy $a/a^\ast = 2.20$ (stable) and $0.50$ (unstable), 
consistent with the sign of $\Re(\lambda)$.

\subsection{Eigenvalue spectra and complex-plane illustration}

For $V(s)=\tanh(s)$, with $c=\tanh(b)$ and $f=\operatorname{sech}^2(b)$, 
the dispersion relation gives
\[
\lambda_\pm(k)=
-\frac{a}{c}\pm\frac{1}{2}\sqrt{\left(\frac{2a}{c}\right)^2
-4\,\frac{2af}{c}\bigl(1-e^{ik}\bigr)}.
\]

We compare
\[
a=2.2\,a^\ast \quad (\text{stable}), 
\qquad
a=0.5\,a^\ast \quad (\text{unstable}).
\]

Stability requires $\Re(\lambda)<0$ for all modes. Crossing into the right half-plane 
produces exponential growth of perturbations
\cite{Strogatz2015,Hirsch2012,Khalil2002}.

\begin{figure}[H]
    \centering
    \includegraphics[width=\textwidth]{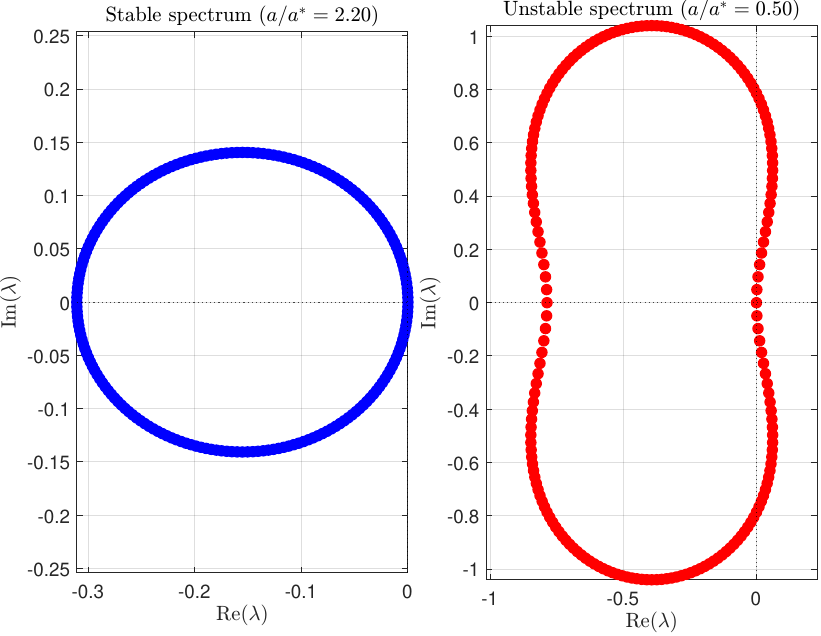}
    \caption{Eigenvalue spectra of the Hybrid OVD model. 
    Left: stable regime ($a>a^\ast$). 
    Right: unstable regime ($a<a^\ast$).}
    \label{fig:hybrid-eigs-complex}
\end{figure}

\section{Numerical Illustration}

We now present numerical simulations of the Hybrid OVD model on a ring
road to verify the analytical stability threshold and to illustrate the
contrast between stable and unstable traffic regimes. The setup follows
the classical protocol of Bando et al.~\cite{Bando1995}: a periodic
system of $N=100$ vehicles is simulated using a forward Euler method
with time step $\Delta t = 0.1$ over a horizon of $T=100$. Two ring
lengths are selected,
\[
L_{\mathrm{stable}} = 200, \qquad L_{\mathrm{unstable}} = 50,
\]
corresponding to equilibrium headways $b = L/N = 2.0$ and $b = 0.5$,
respectively. For each case, the acceleration parameter $a$ is chosen
above or below the theoretical threshold $a^\ast = c\,V'(b)$, giving one
linearly stable simulation and one unstable simulation.

\subsection{Trajectory and velocity plots}

Initial conditions consist of uniform spacing $b=L/N$, together with a
small perturbation applied to the position of the first vehicle. We
record the unwrapped trajectories $x_n(t)$ and velocities $v_n(t)$ for
all vehicles. As in~\cite{Bando1995}, a subset of vehicles is displayed
to highlight the collective behavior.

\begin{figure}[H]
    \centering
    \includegraphics[width=\textwidth]{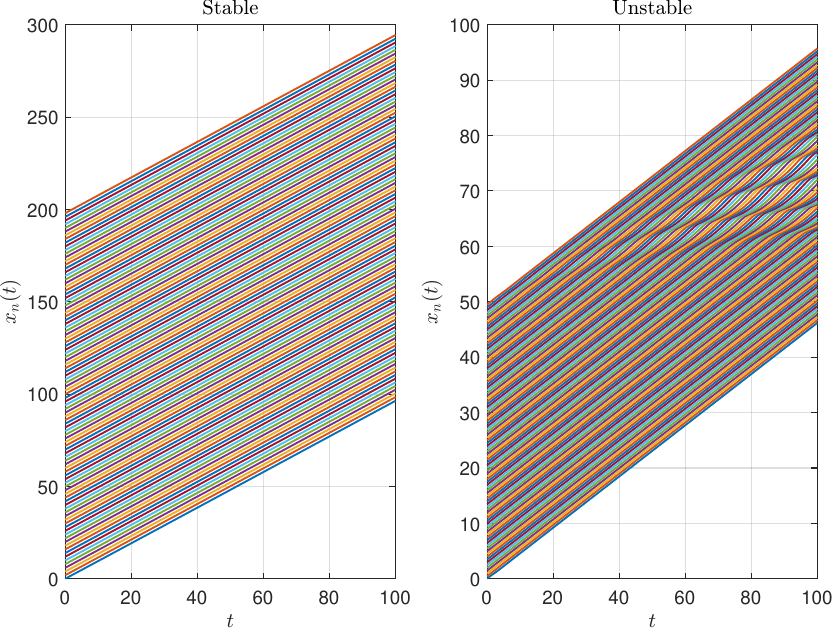}
    \caption{Unwrapped trajectories of all $N=100$ vehicles. 
    \textbf{Left:} stable regime ($L=200$, $b=2.0$, $a/a^\ast=2.20$): 
    perturbations decay and trajectories remain nearly parallel. 
    \textbf{Right:} unstable regime ($L=50$, $b=0.5$, $a/a^\ast=0.50$): 
    perturbations amplify into nonlinear stop-and-go waves. 
    The chosen values of $a/a^\ast$ place the left panel in the 
    stable regime and the right panel in the unstable regime predicted 
    by the stability threshold in Sec.~\ref{sec:stability-longwave}.}

    \label{fig:hybrid-trajectories-allcars}
\end{figure}

Figure~\ref{fig:hybrid-trajectories-allcars} shows that in the stable
case the trajectories remain nearly parallel, and the perturbation
decays. In contrast, when $a < a^\ast$ the disturbance grows and
develops into nonlinear stop-and-go waves, consistent with the
eigenvalue analysis.

For clarity, an extended unstable simulation up to $t=300$ is provided
in~\ref{app:longtraj}, where the persistence and propagation of the
stop-and-go structure are more evident.

\begin{figure}[H]
    \centering
    \includegraphics[width=0.99\textwidth]{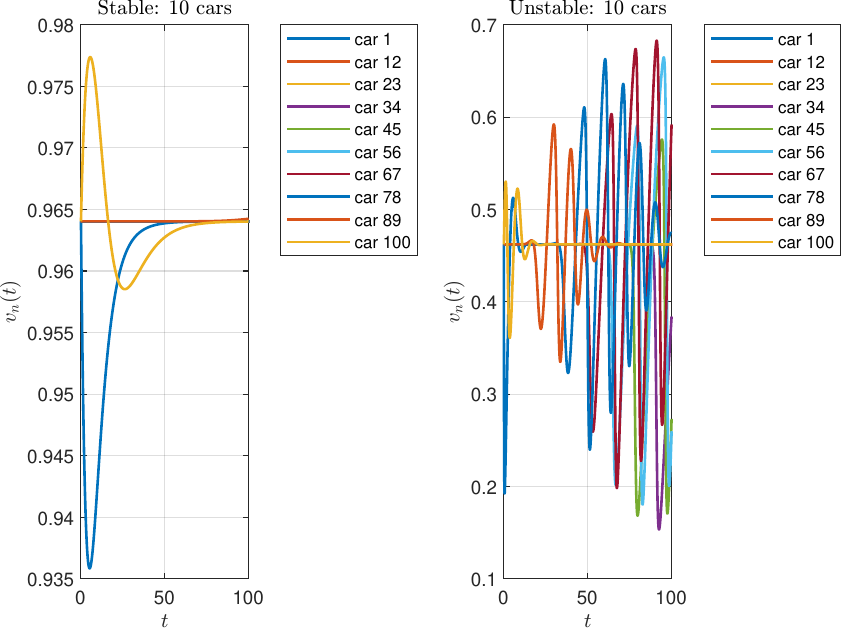}
    \caption{Velocity evolution $v_n(t)$ for selected vehicles under stable 
    (left) and unstable (right) regimes. In the stable case, cars begin near 
    $v_n(0)\approx 0.97$ and remain close to equilibrium. In the unstable case, 
    cars begin at $v_n(0)\approx 0.46$ and develop stop-and-go oscillations. 
    Velocities remain strictly positive, reflecting the smooth form of the 
    $\tanh$ optimal velocity function.}
    \label{fig:velocities_10cars}
\end{figure}

Figure~\ref{fig:velocities_10cars} reports the velocity evolution
$v_n(t)$ for representative vehicles. In the stable case, velocities
remain close to the equilibrium value $v_n(0) \approx 0.97$ (normalized
by $v_{\max}$), with small oscillations that gradually decay. In the
unstable case, although the initial velocities are uniform,
$v_n(0) \approx 0.46$, the perturbation amplifies and produces the
characteristic oscillatory pattern of stop-and-go waves.

A notable feature is that the velocities never drop to zero. This is
consistent with the smooth hyperbolic tangent optimal velocity function
\(
V(\Delta x)=\tanh(\Delta x),
\)
which does not strictly enforce full stops at small headways. As noted
in~\cite{Bando1995}, this behavior is typical of tanh-based OVM
formulations and motivates later variants that introduce cutoffs or
piecewise definitions to allow $V(\Delta x)=0$ under dense congestion
\cite{Helbing2001,Nagatani2002,Jiang2001}.

\subsection{Fourier mode analysis}

To quantify the evolution of perturbations, we compute the Fourier
amplitudes of the deviation from uniform flow. Define
\[
y_n(t) = x_n(t) - \big(nb + c\,t\big),
\]
and
\[
A_k(t) = \left| \sum_{n=0}^{N-1} y_n(t)\,e^{-i \alpha_k n} \right|,
\qquad 
\alpha_k = \frac{2\pi k}{N}.
\]
Following~\cite{Bando1995}, we examine five representative modes
$k = 10,20,30,40,50$.

\begin{figure}[H]
    \centering
    \includegraphics[width=\textwidth]{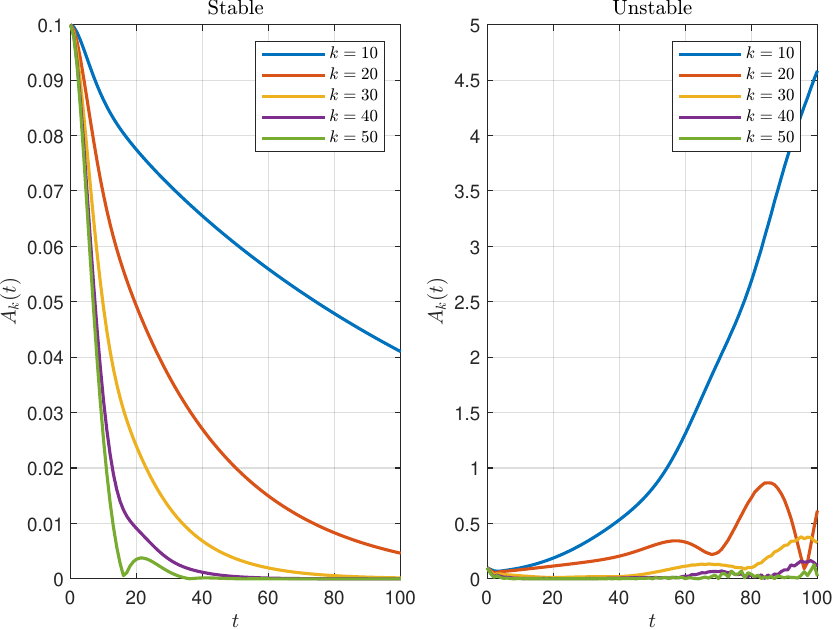}
    \caption{Fourier amplitudes $A_k(t)$ for modes $k=10,20,30,40,50$ in the
    Hybrid OVD model with $N=100$ vehicles. 
    \textbf{Left:} stable regime ($L=200$, $b=2.0$, $a/a^\ast=2.20$): all 
    modes decay, consistent with the prediction $\Re\lambda(k)<0$ for 
    $a>a^\ast$ in Sec.~\ref{sec:stability-longwave}. 
    \textbf{Right:} unstable regime ($L=50$, $b=0.5$, $a/a^\ast=0.50$): 
    selected modes grow, matching the theoretical condition $a<a^\ast$ 
    where long-wave perturbations satisfy $\Re\lambda(k)>0$.}

    \label{fig:hybrid-fourier}
\end{figure}

The results in Fig.~\ref{fig:hybrid-fourier} agree with the dispersion
relation: when $a>a^\ast$, all Fourier modes decay monotonically,
indicating stability, whereas in the unstable regime one or more modes
grow over time, signaling the emergence of stop-and-go behavior.

\subsection{Sensitivity analysis}
\label{sec:sensitivity}

\begin{figure}[h!]
    \centering
    \includegraphics[width=0.95\textwidth]{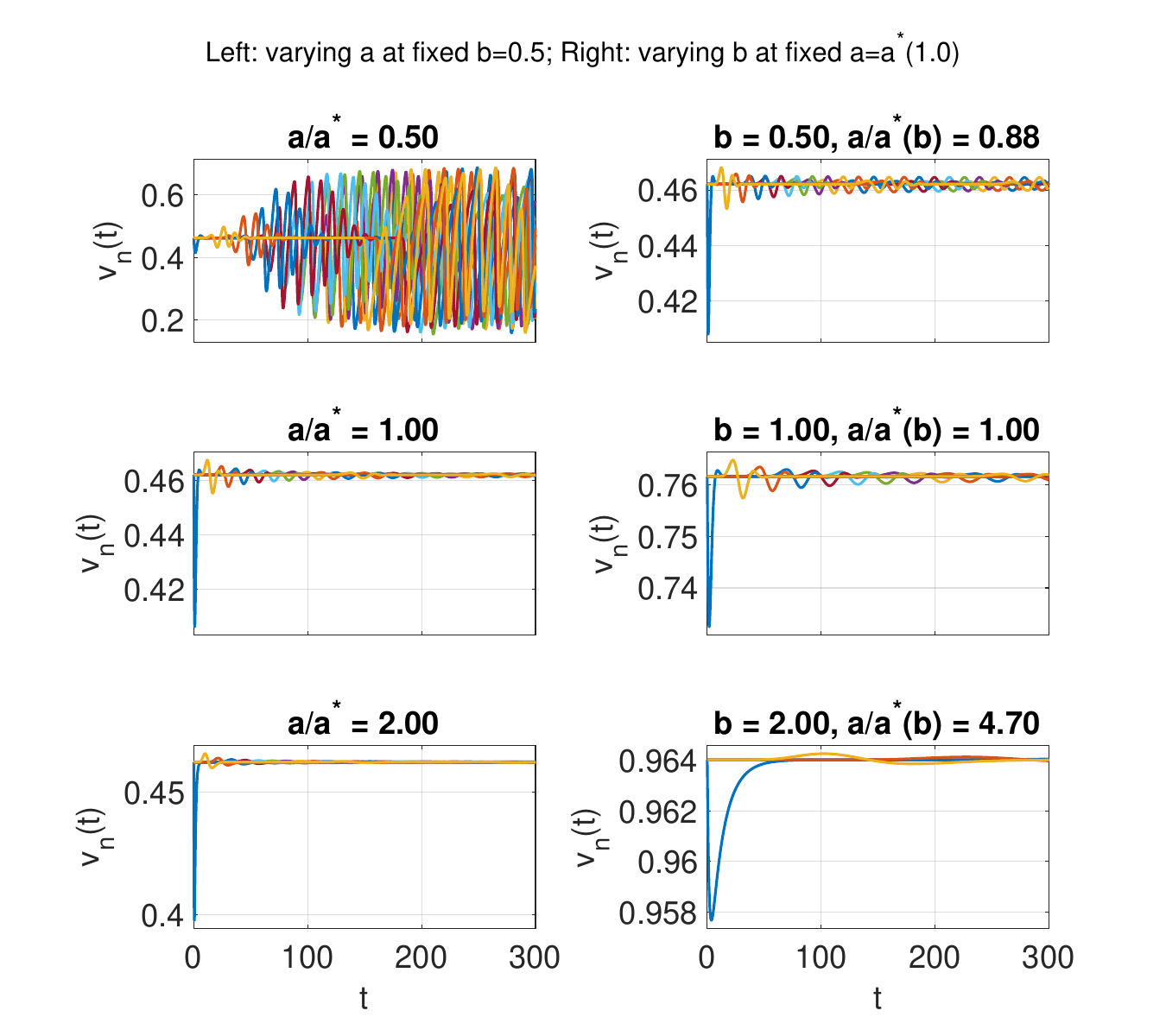}
    \caption{Sensitivity of the Hybrid OVD model. 
    \emph{Left column:} varying the acceleration scale $a$ at fixed headway
    $b=0.5$. 
    \emph{Right column:} varying the headway $b$ at fixed acceleration
    $a=a^*(1.0)$. 
    Each subplot shows the velocity evolution $v_n(t)$ for a subset of vehicles.
    The qualitative behavior in all six cases is consistent with the analytical
    stability threshold $a^*(b)=V(b)V'(b)$.}
    \label{fig:sensitivity}
\end{figure}

The nonlinear dynamics described earlier reveal how uniform traffic flow can
lose stability and develop stop-and-go waves. Since the stability threshold
\[
    a^*(b) = V(b)\,V'(b)
\]
depends on both the acceleration scale $a$ and the headway~$b$, it is important
to examine how the system responds as these parameters vary. To do so, we repeat
the ring-road experiment from Section~5.1, using the same initial perturbation,
but now systematically varying either $a$ (with $b$ fixed) or $b$ (with $a$
fixed). The resulting six simulations are shown together in
Figure~\ref{fig:sensitivity}, arranged in two columns for direct comparison.

\paragraph{Varying the acceleration scale $a$}
The left column of Figure~\ref{fig:sensitivity} shows the effect of changing $a$
while keeping the headway fixed at $b=0.5$. We test three representative
regimes: $a/a^*(b)=0.50$ (below threshold), $1.00$ (critical), and $2.00$
(above threshold). When $a<a^*(b)$, small perturbations grow and form nonlinear
stop-and-go waves. At the threshold $a=a^*(b)$, the oscillations persist with
roughly constant amplitude. When $a>a^*(b)$, all perturbations decay and the
system returns to uniform flow. These behaviors align precisely with the linear
dispersion relation: the sign of $\Re(\lambda)$ changes when $a$ crosses
$a^*(b)$.

\paragraph{Varying the headway $b$}
The right column of Figure~\ref{fig:sensitivity} varies the headway
$b\in\{0.5,1.0,2.0\}$ while keeping the acceleration scale fixed at
$a=a^*(1.0)$, so that $b=1.0$ lies near the stability boundary. Because
$a^*(b)$ increases as density increases (smaller $b$), the effective ratios
$a/a^*(b)$ differ substantially across the three cases. For $b=0.5$, the flow is
close to instability and clear oscillations develop. For $b=1.0$, perturbations
decay slowly, consistent with near-critical behavior. For $b=2.0$, the system is
well inside the stable regime and disturbances damp out quickly.

\paragraph{Summary}
Across both parameter variations, the Hybrid OVD model exhibits behavior that is
fully consistent with the analytical threshold $a^*(b) = V(b)V'(b)$. Instability
appears when $a<a^*(b)$ or when density is high, while strong stability emerges
for $a\gg a^*(b)$ or for large headways. The drag-limited form of the
acceleration ensures that the nonlinear waveforms remain bounded in all cases.

\section{Limitations}
\label{sec:limitations}

The following limitations are not specific flaws of the Hybrid OVD model, but 
rather structural features \emph{inherited from the broader class of 
headway-only car-following laws}.  
They clarify the scope within which the model is intended to operate and 
highlight avenues for natural extensions.

\medskip
\noindent\textbf{(1) Jerk behavior for overspeed states.}
The drag-based saturation produces a smooth and bounded jerk profile for the 
physically relevant regime $0 \le v \le V(\Delta x)$.  
For overspeed states $v>V(\Delta x)$, however, the jerk grows cubically.  
Such overspeed conditions rarely persist in typical traffic, so this behavior 
has negligible effect on realistic simulations.  
It reflects a deliberate modeling choice: the quadratic saturation ensures  
analytical tractability, although more symmetric sigmoidal alternatives 
(e.g.\ $\tanh$, $\arctan$) could remove the cubic overshoot at the cost of 
greater algebraic complexity.

\medskip
\noindent\textbf{(2) Symmetric acceleration--braking response.}
For clarity, the Hybrid OVD model employs a single saturation law for both 
acceleration and deceleration.  
Real vehicles exhibit fundamentally asymmetric dynamics, so adopting separate 
acceleration and braking gains would increase physical realism.  
This modification is straightforward and remains consistent with the overall 
framework.

\medskip
\noindent\textbf{(3) Stability threshold may require high acceleration 
in the absence of velocity-difference feedback.}
Like the classical OVM, the Hybrid OVD model responds only to headway.  
Consequently, the long-wave stability condition 
$a>V(b)V'(b)$ may require accelerations exceeding those achievable by 
passenger vehicles under typical highway conditions.  
This is a well-known limitation of headway-only models, which underestimate 
drivers' responses to closing speeds and therefore rely on artificially strong 
restoring forces to prevent instability.

\medskip
\noindent\textbf{(4) Remedy via anticipatory or velocity-difference terms.}
Introducing a modest anticipatory contribution such as 
$k\,(v_{n+1}-v_n)$—analogous to the Full Velocity Difference Model—can 
substantially lower the stability threshold and restore realistic driver 
sensitivity to closing speed.  
Combining such anticipatory behavior with the Hybrid OVD's physically grounded 
saturation law represents a promising direction for developing a unified 
behavioral--physical car-following framework.

\medskip
Overall, these limitations are \emph{common to headway-only formulations} and 
do not compromise the objectives of the Hybrid OVD model.  
Rather, they position the model as a physically consistent and analytically 
transparent foundation on which richer behavioral and automated-driving 
features can be systematically developed.

\section{Implications and Applications}

The Hybrid OVD model introduced in this work combines two complementary
perspectives on car-following: the behavioral structure of the Optimal
Velocity Model, in which desired speeds depend solely on headway, and the
physical dynamics of drag-limited acceleration. By replacing the classical
linear relaxation law with a nonlinear saturation mechanism, the hybrid model
guarantees bounded accelerations and smooth convergence toward equilibrium.
This modification retains the essential OVM principle that drivers respond
primarily to available spacing, while eliminating the unrealistically large
accelerations and sharp velocity excursions that can arise in the classical
formulation.

From a modeling standpoint, the acceleration parameter $a$ serves as an
interpretable measure of driver aggressiveness or vehicle power, whereas the
headway response $V(\Delta x)$ may be chosen or calibrated to match empirical
observations. This makes the Hybrid OVD model suitable for microscopic
trajectory calibration, controlled experiments, or large-scale numerical
simulations. Because the acceleration law is intrinsically bounded, the model
is numerically robust and physically consistent even in high-density regimes in
which unbounded-relaxation models may exhibit unrealistic dynamics or stability
issues.

\subsection{Connection to prior OVM studies}

Although our theoretical development applies to general optimal velocity
functions, a frequently used choice in the OVM literature is the sigmoid map
\[
    V(\Delta x)
    = \frac{v_{\max}}{2}\!\left[
        \tanh\!\left(\frac{\Delta x-h_c}{\ell}\right) + 1
      \right],
\]
characterized by a free-flow speed $v_{\max}$, a critical headway $h_c$, and a
slope parameter $\ell$. This form has been widely used in classical OVM studies
\cite{Helbing2001,Treiber2013} because it combines analytical convenience with
a realistic saturation toward free-flow speed. In our numerical demonstrations,
we adopt a simplified hyperbolic tangent for clarity, but the Hybrid OVD
framework accommodates the sigmoid law without altering its analytical
structure. The parameters $h_c$ and $\ell$ then directly control the steepness
of the headway--speed response and therefore influence the stability threshold
$a^*(b)=V(b)V'(b)$.

In this sense, the Hybrid OVD model may be viewed as a natural extension of the
classical OVM: the widely used optimal velocity function is retained, but the
acceleration rule is replaced by a more realistic drag-inspired dynamics that
ensures bounded and physically plausible behavior.

\subsection{Applications}

The structure of the Hybrid OVD model enables several applications across
traffic flow modeling and dynamical systems analysis.  
First, the model provides a tractable framework for exploring nonlinear wave
formation and stability transitions in car-following systems. The explicit
stability threshold identified in this work clarifies how wave onset depends on
density (through $b$) and acceleration capability (through $a$).

Second, the explicit acceleration bound improves robustness for microscopic
simulations, particularly in high-density scenarios where unbounded-relaxation
models may produce unrealistic or numerically unstable trajectories. This makes
the hybrid formulation suitable for integration into more complex models, such
as multi-lane, network-level, or mixed-traffic simulations.

Third, the parameter $a$ naturally captures heterogeneity across drivers or
vehicles, allowing the model to represent a spectrum of behaviors from cautious
to aggressive. This feature is also relevant for automated or assisted driving
contexts, where bounded acceleration and smooth convergence are important design
criteria for control algorithms.

Fourth, the Hybrid OVD model is compatible with established calibration and
assessment frameworks for microscopic trajectory data. In particular, datasets
such as NGSIM could be used to estimate parameters and evaluate model fidelity
using procedures developed by Punzo, Montanino, and Ciuffo \cite{Punzo2016},
who demonstrated systematic methods for calibrating and validating car-following
laws against real vehicle trajectories. Incorporating such calibration techniques
provides a pathway for empirical grounding and future data-driven refinement of
the hybrid model.

\medskip
\noindent
These findings clarify how physical acceleration limits influence nonlinear 
wave behavior and motivate several extensions discussed below.

\medskip
\noindent
\textbf{Broader transportation implications.}
Beyond its analytical contributions, the Hybrid OVD model offers a physically
grounded framework for evaluating stability and driver-comfort constraints in
transportation operations. Bounded acceleration and realistic convergence
behavior are directly relevant to energy use, environmental impact, and
automated-vehicle control algorithms. As a result, the model can support the
development of safer and smoother traffic-management strategies, making it
useful not only for theoretical studies but also for planning and operational
analysis in modern transportation systems.

Overall, the Hybrid OVD model provides a bridge between behavioral and physical
approaches to car-following. By enriching the classical OVM with a simple
drag-based saturation mechanism, it combines interpretability, analytical
tractability, and physical realism, yielding a versatile tool for both
theoretical study and numerical exploration.

\subsection{Future Directions}
\label{sec:futuredirections}

Although the Hybrid OVD model enforces physically realistic acceleration
bounds, its current formulation does not incorporate driver anticipation or
velocity-difference feedback, features known to improve stability in classical
extensions of the OVM. A natural next step is to augment the Hybrid OVD
equation with a term proportional to the relative speed between successive
vehicles:
\begin{equation}
    \frac{dv_n}{dt}
    = a\!\left(1-\frac{v_n^2}{V(\Delta x_n)^2}\right)
    + k\,(v_{n+1}-v_n),
\end{equation}
where $k>0$ measures the strength of anticipatory response. Linearization about
the uniform state $(v=V(b))$ yields the modified long-wave stability condition
\[
    a > V(b)\,[V'(b)-k],
\]
showing that even modest anticipation ($k>0$) can reduce the required
acceleration scale for stability. Future work will examine the nonlinear wave
behavior of this hybrid--anticipatory model, explore its parameter regimes, and
compare its stability properties with those of classical FVDM-type formulations
under the same physical constraints.

\section{Conclusion}

We have introduced a Hybrid Optimal Velocity--Drag (OVD) model that
enhances the classical OVM by embedding a physically motivated,
drag-based saturation law into the headway-dependent desired-speed
framework. This modification addresses a key deficiency of the classical
OVM: the possibility of unrealistically large accelerations when large
speed discrepancies occur. The Hybrid OVD model enforces bounded,
smoothly decaying accelerations while preserving the nonlinear
instabilities responsible for the formation of stop-and-go waves.

Analytically, we derived the corresponding dispersion relation and
identified the long-wave stability threshold, demonstrating that the
hybrid formulation retains the essential structure of the OVM while
rescaling the local relaxation dynamics. Numerically, simulations on a
ring road illustrate that the hybrid model reproduces both stable flow
and the nonlinear growth of stop-and-go oscillations, with acceleration
and velocity trajectories that remain physically plausible. Together,
these results show that the Hybrid OVD framework achieves a balance of
behavioral interpretability and dynamical realism.

\rev{Unlike existing extensions that primarily modify interaction laws or incorporate control and delay mechanisms, the proposed formulation isolates the role of physically grounded acceleration saturation and demonstrates how this mechanism alone reshapes the stability structure of traffic flow. This provides a new mechanistic perspective that complements established behavioral and control-based approaches.}

From a modeling perspective, the acceleration parameter \(a\) offers an
intuitive representation of driver aggressiveness or vehicle power,
while the headway response \(V(\Delta x)\) remains compatible with
standard OVM calibrations. This makes the hybrid model suitable for
empirical trajectory fitting and for use in large-scale microscopic
simulations, including settings involving both human-driven and automated
vehicles. Because the formulation is compact and analytically
tractable, it provides a promising foundation for extensions such as
velocity-difference feedback, anticipatory behavior, asymmetric braking
limits, and automated-vehicle control strategies.

In summary, the Hybrid OVD model strengthens the link between behavioral
car-following theory and physical vehicle dynamics. By combining bounded,
drag-limited acceleration with the familiar OVM structure, it provides a
practical and interpretable tool for studying traffic instabilities and
for developing next-generation traffic-flow models and control systems. 
\rev{More broadly, the framework highlights how incorporating physically consistent acceleration constraints can influence stability, wave formation, and control design in modern traffic systems.}
In addition, the model’s physically grounded acceleration constraints and 
stability properties offer insights that are relevant to traffic operations, 
automated-vehicle design, and the development of smoother and safer mobility 
systems.





\appendix
\section{Extended Unstable Trajectories}
\label{app:longtraj}

To further illustrate the nonlinear growth of stop-and-go oscillations,
we extend the unstable simulation horizon to $t=300$. The longer time
window makes the persistence and propagation of traffic jams more
visible, revealing the characteristic travelling-wave structure that
emerges once perturbations leave the linear regime.

\begin{figure}[H]
    \centering
    \includegraphics[width=\textwidth]{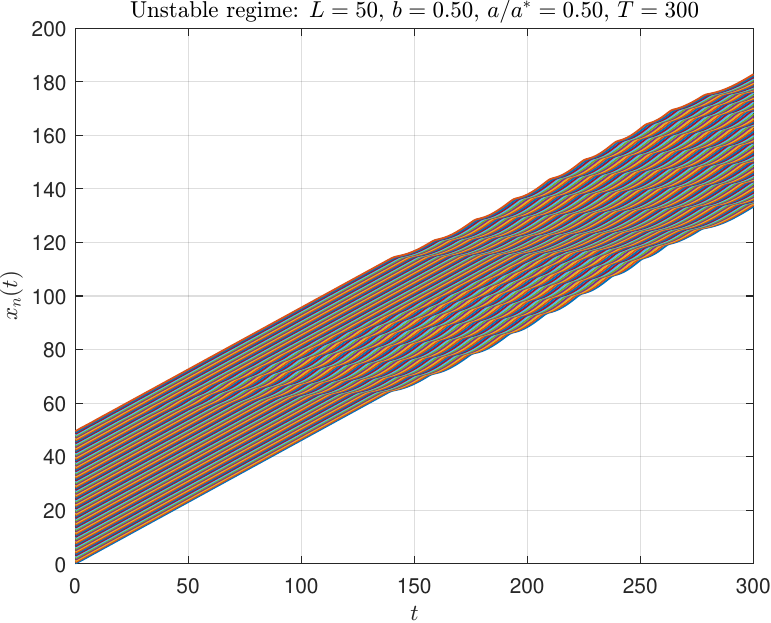}
    \caption{Extended unstable simulation of the Hybrid OVD model
    with $N=100$ vehicles, ring length $L=50$ ($b=0.5$), and
    $a/a^\ast=0.50$. Extending the simulation to $T=300$ highlights
    the nonlinear amplification of perturbations into persistent,
    travelling stop-and-go waves.}
    \label{fig:longtraj}
\end{figure}

\section{Jerk in the OVM and Hybrid OVD Model}
\label{app:jerk-analysis}

We compute the \emph{jerk}, defined as the time derivative of
acceleration,
\[
j = \frac{d^2 v}{dt^2},
\]
for both the classical OVM and the Hybrid OVD model.  For clarity,
we fix the desired velocity at a constant value $V$, allowing the
internal structure of each model to be compared directly.

\paragraph{Classical OVM}

The classical OVM uses the linear relaxation law
\[
\dot v = \alpha (V - v),
\]
so differentiating once more gives
\[
j_{\mathrm{OVM}}
   = \frac{d}{dt}\big[\alpha(V-v)\big]
   = -\alpha^2 (V - v).
\]
Thus the jerk is linear in the deviation from $V$: negative when
$v<V$, positive when $v>V$, and unbounded as $|v-V|$ grows.

\paragraph{Hybrid OVD model}

For the Hybrid OVD model, the acceleration law
\[
\dot v = a\left(1-\frac{v^2}{V^2}\right)
\]
yields
\[
j_{\mathrm{Hybrid}}
= \frac{d}{dt}\left[a\Bigl(1-\frac{v^2}{V^2}\Bigr)\right]
= -\frac{2a}{V^2}v\,\dot v
= -\frac{2a^2}{V^2}\,v\left(1-\frac{v^2}{V^2}\right).
\]
This jerk is cubic in $v$: it vanishes at $v=0$ and $v=V$, is negative
for $0<v<V$, and becomes positive for $v>V$.

\paragraph{Discussion}

Near equilibrium, both models behave similarly: the sign of jerk
indicates whether $v$ is above or below $V$.  Globally, however,
their behaviors differ markedly.  
The OVM jerk grows linearly and without bound as $|v-V|$ increases,
whereas the Hybrid jerk is bounded for $0\le v\le V$ and only grows
under overspeed ($v>V$).  

A full comparison of the jerk laws, including plots, is now presented
in the main text (Fig.~\ref{fig:ovm-hybrid-jerk}; Sec.~\ref{sec:local}).  
Here we note only that the overspeed cubic growth is a consequence of
the symmetric quadratic saturation law; its implications and possible
remedies are discussed in Sec.~\ref{sec:limitations}.




\bibliographystyle{IEEEtran}
\bibliography{sample_revised}

\end{document}